\documentclass[prc,twocolumn,letterpaper,showpacs,amsmath,amssymb,floatfix,nofootinbib,superscriptaddress]{revtex4}
\pdfoutput=1
\usepackage{graphicx}
\usepackage{bm,url,braket}
\usepackage{xcolor}
\usepackage{amsmath}

\begin{document}

\title{Fully self-consistent calculations of nuclear Schiff moments}

\author{Shufang Ban}

\affiliation{Dept.\ of Physics and Astronomy, University of North Carolina,
Chapel Hill, NC, 27516-3255, USA}
 \affiliation{Nuclear Theory Center and Department of Physics, Indiana
 University, 701 E. Third St., Bloomington, IN, 47405, USA}

\author{Jacek Dobaczewski}
\affiliation{Institute of Theoretical Physics, Warsaw University ul. Hoza 69,
PL-00681 Warsaw, Poland}
\affiliation{Department of Physics, P.O. Box 35 (YFL), FI-40014 University of
Jyvaskyla, Finland}

\author{Jonathan Engel}
\affiliation{Dept.\ of Physics and Astronomy, University of North Carolina,
Chapel Hill, NC, 27516-3255, USA}

\author{A.\ Shukla}
\affiliation{Dept.\ of Physics and Astronomy, University of North Carolina,
Chapel Hill, NC, 27516-3255, USA}
\affiliation{Rajiv Gandhi Institute of Petroleum Technology, Ratapur Chowk, Raebareli -229316, U.P., India}

\begin{abstract}
We calculate the Schiff moments of the nuclei $^{199}$Hg and $^{211}$Ra in
completely self-consistent odd-nucleus mean-field theory by modifying the
Hartree-Fock-Bogoliubov code \texttt{HFODD}.  We allow for arbitrary shape
deformation, and include the effects of nucleon dipole moments alongside those
of a CP-violating pion-exchange nucleon-nucleon interaction.  The results for
$^{199}$Hg differ significantly from those of previous calculations when the
CP-violating interaction is of isovector character.
\end{abstract}


\maketitle
\section{Introduction}\label{sec:intro}

There are compelling reasons to believe in a source of measurable CP violation
from outside the Standard Model of particle physics.  Supersymmetry and other
theories that lessen the hierarchy problem\footnote{See. e.g., Ref.\
\cite{zee03} for a short but clear statement of the problem.} typically
introduce many new fields with CP-violating phases into Lagrangians.  If
unsuppressed, such phases should be observable, now or in the foreseeable
future.  And they are needed if the imbalance between matter and antimatter in
the universe is the result of a CP asymmetry; in the Standard Model CP
violation is too weak to be responsible \cite{trodden99}.

As long as the CPT theorem holds, one can search for time-reversal (T)
violation in lieu of CP violation.  One of the best ways to observe T
violation from beyond the Standard Model is by measuring nonzero static
electric dipole moments (EDMs) in systems with nondegenerate ground states.
Standard-Model CP violation is suppressed in flavor-conserving processes, so an
observed EDM anywhere near current limits would imply new physics.
Experimental groups have been steadily lowering the upper limits on EDMs to the
point that one might reasonably expect an observation in the near future. (Much
of supersymmetry parameter space has already been covered.)  For now, the
nonobservation of an EDM in the diamagnetic atom $^{199}$Hg places tight upper
limits on CP violation, and measurements in other diamagnetic systems ---
$^{129}$Xe, $^{223,225}$Ra, and $^{223}$Rn --- may soon do even better. 

Whether these experiments eventually see a nonzero EDM or just continue to set
limits, their interpretation requires us to understand the dependence of atomic
EDMs on the strength of CP violation at the fundamental level.  Doing so
involves calculations at several scales.  QCD determines the dependence on
fundamental physics of the neutron EDM and related quantities such as effective
P- and T-violating meson-nucleon coupling constants.  Nuclear physics then
translates these quantities into P- and T-violating nuclear moments, which
in turn contribute to atomic EDMs.

The role of nuclear physics in this chain is more subtle than it appears at
first glance because the atomic electrons screen nuclear EDMs \cite{schiff63}.
As a result, the nuclear quantity that plays the largest role in inducing
atomic EDMs is not the nuclear dipole moment, but rather the ``Schiff moment'',
\begin{equation}
\label{eq:exval}
S \equiv \bra{0} S_z \ket{0}_{M=J} \,,
\end{equation}
that is, the ground state expectation value (in the substate that is fully
polarized along the $z$ axis) of the $z$-component of the one-body ``Schiff
operator.'' This vector operator is given approximately by
\begin{equation}
\label{eq:op}
\bm{S} = \bm{S}^\text{ch} + \bm{S}^\text{nucleon}  \,,
\end{equation}
where
\begin{align}
\label{eq:opch}
\bm{S}^\text{ch}&=\frac{e}{10}\,\sum_{p=1}^Z\left(r^2_p-\frac{5}{3}\,
\braket{r^2}_\text{ch}\right) \bm{r}_p \,. \\
\label{eq:opnucl}
\bm{S}^\text{nucleon} & = \frac{1}{6} \sum_{j=1}^A \bm{d}_j (r_j^2 -
\braket{r^2}_\text{ch}) \nonumber \\
&+\frac{1}{5} \sum_{j=1}^A \left(\bm{r}_j(\bm{r}_j
\cdot \bm{d}_j) - \frac{r^2_j}{3} \bm{d}_j \right) + \ldots
\end{align}
Here $e$ is the charge of the proton, $\langle r^2\rangle_{\rm ch}$ is the mean
squared radius of the nuclear charge distribution, $\bm{d}_j$ is the EDM of
nucleon $j$, and the omitted terms in Eq.\ (\ref{eq:opnucl}) are smaller than
those included by about the square of the ratio of the proton radius to the
nuclear radius.  The sum in Eq.\ (\ref{eq:opnucl}) is over all nucleons, while
that in Eq.  (\ref{eq:opch}) is restricted to protons.

The two terms in Eq.\ (\ref{eq:op}) reflect the two ways in which a nucleus can
acquire Schiff moments.  A P- and T-violating nucleon-nucleon interaction
generates a corresponding charge distribution and a contribution to $S$ from
the operator $\bm{S}^\text{ch}$ in Eq.\ (\ref{eq:opch}), while nucleon EDMs
generate a contribution from $\bm{S}^\text{nucleon}$ in Eq.\
(\ref{eq:opnucl}).  Both contributions can be induced by effective P- and
T-violating pion-nucleon coupling constants:  a pion-exchange graph with one
such coupling generates the effective nucleon-nucleon interaction and a pion
loop graph with one generates nucleon EDMs.  The two-body interaction is
\begin{align}
   V_{PT}= {} &\frac{{\rm g}}{8\pi m_{\rm N}} \sum_{i<j}
   \bigg\{\Big[{\rm \bar{g}}_0\left(\bm{\tau}_i\cdot\bm{\tau}_j\right)
   -\frac{{\rm \bar{g}}_1}{2}(\tau_i^z+\tau_j^z) \\
   &\hspace*{-0.3cm}+ {\rm \bar{g}}_2(3\tau_i^z\tau_j^z-
   \bm{\tau}_i\cdot\bm{\tau}_j)
   \Big]\left(\bm{\sigma}_i-\bm{\sigma}_j\right)
   \nonumber \\
   &\hspace*{-0.3cm} - \frac{{\rm \bar{g}}_1}{2}
   (\tau_i^z-\tau_j^z) (\bm{\sigma}_i+\bm{\sigma}_j)\bigg\}\cdot
   \bm{\nabla}_i \frac{{\rm exp}\left(-m_\pi|\bm{r}_i-\bm{r}_j|\right)}
   {|\bm{r}_i-\bm{r}_j|} \,, \nonumber
\end{align}
and the nucleon EDM operator (for nucleon $j$), in the leading chiral
approximation\footnote{Nucleons can get EDMs in other ways, e.g.\ from quark
EDMs \cite{pospelov05}, but we assume here for simplicity that the $\bar{\rm
g}$'s are the only relevant low-energy CP-violating parameters.  The dependence
of nuclear Schiff moments on the nucleon EDMs, no matter what their source, can
be extracted from our analysis by dividing out the $\bar{\rm g}$-dependent
prefactor in Eq.\ (\ref{eq:dn}).} is
\begin{equation}
\label{eq:dn}
\bm{d}_j = \frac{e\rm{g}}{4\pi^2 m_N} {\rm ln}\frac{m_N}{m_\pi} (\bar{\rm
g}_0-\bar{\rm g}_2) \bm{\sigma}_j \tau^z_j\,,
\end{equation}
In these two equations $\hbar=c=1$, $m_\pi$ is the mass of the pion, $m_{\rm
N}$ is that of the nucleon, $\tau^z$ gives +1 when acting on a neutron, ${\rm g}
\equiv 13.5$ is the strong $\pi$\text{NN} coupling constant, and the $\bar{\rm
g}_i$ are dimensionless isoscalar ($i=0$), isovector ($i=1$), and isotensor
($i=2$) P- and T-violating $\pi$\text{NN} coupling constants.  These last
quantities depend on the unknown fundamental source of CP violation, and so are
primitive in our treatment.  A QCD calculation can in principle relate them to
quantities in extra-Standard-Model theories.

Since $V_{PT}$ is extremely weak and the nucleon EDM in Eq.\ (\ref{eq:dn}) is
extremely small, the Schiff moment, to very high accuracy, is linear in the
$\pi\text{NN}$ couplings $\bar{\rm g}_i$.  We write it as
\begin{equation}
\label{eq:a}
S=(a_0 +b)\, {\rm g \bar{g}}_0 + a_1\, {\rm g \bar{g}}_1 + (a_2-b)\, {\rm g \bar{g}}_2 \,.
\end{equation}
The $a_i$ specify the dependence of $S$ on the P- and T-violating
interaction $V_{PT}$, and $b$ specifies its dependence on the nucleon dipole
moments $\bm{d}_j$.  All relevant nuclear structure information is encoded in
these coefficients.

The $a_i$ have been calculated before, with varying degrees of sophistication,
in nuclei used in or considered for experiments.  Except in a few nuclei with
strong octupole deformation, all prior has proceeded in two steps:  some kind
of mean-field calculation in which the polarizing effects of the last (valence)
nucleon were neglected, followed by an explicit treatment of the correlations
induced by the interaction of the valence nucleon with the rest.  Ref.\
\cite{flambaum86}, the first such calculation, used a phenomenological
Wood-Saxon potential as the mean field and allowed the valence-core interaction
to excite only non-collective one-particle one-hole configurations.  Refs.\
\cite{dmitriev03} and \cite{dmitriev05} obtained the mean field through an
approximate Hartree-Fock calculation and used a simple residual strong
interaction and linear-response theory (that is, the random phase approximation
(RPA)) to include collective corrections to the simple excitations considered
in Ref.\ \cite{flambaum86}.  Finally, Ref.\ \cite{jesus05} carried out a
self-consistent Skyrme-interaction-based calculation to obtain the mean field,
and followed that with a diagrammatic treatment (with the same Skyrme
interaction) of most but not all of the quasiparticle-RPA (QRPA) response
generated by the valence-core interaction.

In the work reported here, we modify the Hartree-Fock-Bogoliubov code
\texttt{HFODD} \cite{dobaczewski09} to carry out completely self-consistent
mean-field calculations directly in the nuclei of interest.  That is, we
\begin{enumerate}
\item treat $V_{PT}$ on the same footing as the strong interaction,
\item treat all the nucleons, including the last, on the same footing in
mean-field theory.
\end{enumerate}
These steps make our treatment essentially equivalent to a fully
self-consistent treatment of the even nucleus followed by the self-consistent
inclusion of all linear-response collectivity induced by the valence-core
interaction.  Thus, unlike the work of Refs.\ \cite{dmitriev03,dmitriev05} our
calculation is completely self consistent, and unlike the work of Ref.\
\cite{jesus05} it includes all core-polarization effects, in a unified way to
boot.  In addition, our mean-field can (and often will) be deformed.  All prior
calculations in systems without octupole deformation assumed spherical ground
states.  In nuclei such as $^{199}$Hg the quadrupole deformation may well be
large enough to affect Schiff moments; the successful M\"{oller}-Nix
phenomenology \cite{moller95} predicts deformation parameters $\beta_2=-0.122$
and $\beta_4= -0.032$, values that are hardly negligible.
Finally, we project our states onto those with well-defined angular momentum
(after variation), going beyond the usual rigid-rotor approximation.  This step
is essential in nuclei that are only weakly deformed.


\section{Method and Tests}
\label{sec:method}

We begin with a more precise statement of the relation between the perturbative
treatment of interactions within linear-response theory, that is, the RPA or
QRPA, and a non-perturbative treatment in mean-field theory.  Consider, for
example, an even-even nucleus with $Z+N=A$ nucleons, neglecting pairing
temporarily to simplify the situation.  It is not hard to show
\cite{brown70,blaizot86} that the one-body density matrix obtained from a
Hartree-Fock (HF) calculation in the neighboring odd nucleus with one more
neutron is related to that obtained from a corresponding (much easier)
calculation in the even nucleus by:
\begin{equation}
\label{eq:rhodivide}
\rho^{A+1}_{a,b} = \rho^{A}_{a,b} + \rho^v_{a,b} + \sum_{c,d}
R^{A}_{ab,cd} \,
h^v_{cd} + \ldots
\end{equation}
where $\rho^{A+1}$ and $\rho^{A}$ are density matrices (isoscalar or isovector)
for the odd and even nuclei, $\rho^v$ is the density matrix associated with the
valence neutron in the first empty orbit produced by the even-nucleus mean
field, $h^v$ is the additional mean-field Hamiltonian created by that valence
nucleon, $R^{A}$ is the zero-frequency RPA response function for the even-even
core, and the neglected terms are higher order in $h^v$.  Eq.\
(\ref{eq:rhodivide}) generalizes predictably when pairing is included via
Hartree-Fock-Bogoliubov (HFB) theory and the QRPA.  When the interaction
$V_{PT}$ is included, it affects Schiff moments only through the last term.

We can now more precisely characterize previous calculations, which were based
on approximate representations of the right-hand side of Eq.\
(\ref{eq:rhodivide}).  Refs.\ \cite{dmitriev03,dmitriev05} used a simple
Landau-Migdal strong interaction and approximate self consistency in
determining the densities and RPA response function $R^{A}$, and treated
$V_{PT}$ without further approximation.  Ref.\ \cite{jesus05} used full-fledged
Skyrme interactions and retained self consistency everywhere, but obtained the
response function $R_A$ by first neglecting $V_{PT}$, then adding first-order
corrections through a series of diagrams, some of which were omitted.  Both
calculations imposed spherical symmetry everywhere.  Here we calculate the
left-hand side of Eq.\ (\ref{eq:rhodivide}) directly in mean field theory,
without the intermediary of response functions and with no approximations or
imposed symmetries.  Of course, the Skyrme interactions we use are not perfect,
but they are the current state of the art.

Mean-field calculations in odd nuclei are notoriously tricky \cite{schunck10}.
Because the valence nucleon can polarize the rest, odd systems are more likely
than their even-even neighbors to have complicated triaxial shapes; $^{129}$Xe,
which has a tight limit on its atomic EDM, is an example.  Unless one projects
triaxial intrinsic states onto states with good angular momentum before the
mean-field variation, it is difficult to ensure that the component with the
correct angular momentum is a significant part of the wave function.  Moreover,
triaxial systems are often soft, meaning that the wave function corresponding
to the absolute minimum energy may not more significantly represent the nuclear
state than other wave functions with only slightly higher energies.  We
therefore will restrict ourselves to axially symmetric systems in which the
spin aligns along the symmetry axis; in such states we can ensure a significant
component with a given $J$ by selecting states for which the intrinsic angular
momentum z-projection $K$ is equal to $J$.  We sometimes pay the price that the
desirable configurations are not the lowest ones and that are solutions are
marginally unstable; we discuss those difficulties below.

To implement our procedure, we employ a modified version of the
state-of-the-art code \texttt{HFODD} (see Ref.\ \cite{dobaczewski09} and
references therein), which uses a symmetry-unrestricted three-dimensional
harmonic-oscillator (HO) basis to carry out Skyrme HF or HFB calculations.  Our
modification is to add $V_{PT}$ to the Skyrme interaction, allowing the
calculation of Schiff moments.  An initial step, reported in Ref.\
\cite{dobaczewski05}, was to represent $V_{PT}$ as a sum of Gaussians in order
to ease calculation in the HO basis, and evaluate its expectation value at the
end of the calculation in octupole-deformed nuclei.  Here we extend that scheme
and incorporate it into the self-consistent loop; the code evaluates the
expectation value of $V_{PT}$ and the corresponding mean fields, which are the
new ingredient, at every iteration. (We have actually coded the mean fields
only in the normal particle-hole mean channel; we deal with the pairing field
through a trick discussed below.) It then adds the P- and T-violating mean
fields to those coming from the Skryme interaction, so that all forces are
treated in the same way.  The resulting P- and T-violating polarization
produces a nonzero expectation value for the Schiff operator $S_z^\text{ch}$ in
Eq.\ (\ref{eq:opch}).  To calculate the expectation value of
$S_z^\text{nucleon}$, we simply use the HF or HFB wave functions obtained
without the addition of $V_{PT}$.

To check the results, we also incorporate the direct part of $V_{PT}$ in a
completely different way.  The direct P- and T-violating mean field can be
written\footnote{The term containing $\bm{s}_1$ was omitted in Ref.\
\cite{engel03}} as
\begin{align}
\label{eq:wdir}
v_{PT}^d &=\frac{\rm g}{8\pi m_N} \int d\bm{r}' \,
\frac{e^{-m_{\pi} |\bm{r}-\bm{r}'|}}{|\bm{r}-\bm{r}'|} \\
& \times \Big\{\left[\bar{\rm g}_1 -(\bar{\rm g}_0+2\bar{\rm g}_2)\tau_{z}\right]
\bm{\nabla} \cdot \bm{s}_1(\bm{r}') \nonumber \\
& \hspace{1.5em} + \bm{\sigma} \tau_{z} \cdot \left\{(\bar{\rm g}_0+
2\bar{\rm g}_2)\bm{\nabla}
\rho_1(\bm{r}') - \bar{\rm g}_1 \bm{\nabla} \rho_0(\bm{r}')
\right\}\Big\} \,, \nonumber
\end{align}
where $\bm{s}_1$ is the isovector spin density (see the appendix of Ref.\
\cite{bender02}, where the density is called $\bm{s}_{10}$, for the exact
definition), and $\rho_0$, $\rho_1$ are the usual isoscalar and isovector
number densities.  This representation as the folding of a Yukawa function with
a source density is similar to the representation of the Coulomb potential as
the folding of the function $1/|\bm{r}-\bm{r}'|$ with the charge density.  We
therefore adapt the existing Green-function-based routine for calculating the
direct Coulomb potential in \texttt{HFODD} to the evaluation of the direct P-
and T-violating mean field $v_{PT}^d$.

Finally, to further check the self-consistent solution, we note that before
projection in an axially symmetric nucleus, one should obtain the same Schiff
moment to leading order in an arbitrary constant $\lambda$ by
\begin{itemize}
\item[a)] Solving self-consistent field equations with $H \equiv
H_\text{Skyrme} + \lambda V_{PT}$, and then evaluating the expectation value of
$S^\text{ch}_z/\lambda$,
\item[b)] Solving the mean-field equations with $H \equiv H_\text{Skyrme} +
\lambda S^\text{ch}_z$ and then evaluating the expectation value of
$V_{PT}/\lambda$.
\item[c)] Solving the mean-field equations with $H \equiv H_\text{Skyrme}$ and
then evaluating
\begin{equation}
\label{eq:pert}
\sum_i \frac{\bra{0} S^\text{ch}_z \ket{i}_\text{RPA} \bra{i} V_{PT} \ket{0}_\text{RPA}}
{(E_0-E_i)} + c.c \,,
\end{equation}
\end{itemize}
where the subscripts on the kets mean that the transition matrix elements are
evaluated in RPA (or QRPA). Procedure a) above defines the problem we're tyring
to solve.  Procedure b) serves as a check and, moreover, is our primary
procedure in nuclei with pairing.  The reason, as mentioned above, is that
although we can evaluate the expectation value of $V_{PT}$ (including the
pairing parts), we cannot evaluate the corresponding pairing field, so that we
cannot include all the effects of pairing in procedure a).  Finally, regarding
the RPA or QRPA: although we cannot do an RPA or QRPA calculation in a deformed
or odd-A nucleus, we can use procedure c) as a test in a spherical nucleus.  A
full odd-A QRPA evaluation, even there, would involve adding all the
complicated diagrams in Ref.\ \cite{jesus05}, so we make our test in the
approximation that last nucleon feels the strong mean-field from the other
nucleons but acts on them in turn only weakly (through $V_{PT}$). This makes it
sufficient to apply the QRPA to the even-even core.

To implement this ``weak-valence-field'' approximation, in a closed-shell+1
nucleus such as $^{57}$Ni, we first calculate the self-consistent ground-state
in the even-even neighbor $^{56}$Ni without including $V_{PT}$ in the
Hamiltonian, and then allow the valence neutron to occupy the first empty
neutron orbit.  We then calculate the P- and T-violating mean field that
that neutron produces (restricting ourselves for simplicity to the dominant
direct part) by evaluating its contribution to $v_{PT}^d$ in Eq.\
(\ref{eq:wdir}).  We then use this mean field as an external P- and
T-violating source for the $^{56}$Ni core.  The Schiff moment of $^{57}$Ni in
the weak-valence-field approximation is then the moment of the $A=56$ core
induced by the external source.

We can implement the procedure in mean-field theory by adding the external
source $v_{PT}^d$ or $S_z^{\text{ch}}$ for $^{56}$Ni to $H_\text{Skyrme}$ as in
procedure a) or b) above, or in the RPA by substituting $v_{PT}^d$ for $V_{PT}$
in procedure c). The first two routes are straightforward and give identical
results but the spherical RPA requires a decomposition of $v_{PT}^d$ into
spherical multipoles.  To make that simpler, we use the zero-range (infinite
pion-mass) approximation, which reduces the Yukawa function in Eq.\
(\ref{eq:wdir}) to a delta function, when carrying out any of the three
procedures a), b), and c).  Even so, we can always expect slight differences
between the results of procedure c) and the others because of slight
differences in the single-particle spaces underlying the mean-field and RPA
calculations.  In the former, we include single-particle HO basis states with
up to 22 $\hbar \omega$ of excitation energy.  In the latter, which we carry
out with the spherical HFB code HFBRAD \cite{[Ben05]} and the QRPA code QRPAsph
\cite{terasaki2006}, we include single-particle spherical-box states with
energies up to 100 MeV.  Despite the single-particle differences, the results
of the procedures a) and c), displayed in Tab.\ \ref{t:table1} for the Skyrme
interaction SKM$^*$ \cite{[Bar82]}, are extremely close.

The table also compares the results of procedures\footnote{In this nucleus, an
accurate mean-field result requires dealing with the center-of-mass shift that
results from the fixed external source $v^d_{PT}$; the task is easier in
procedure b) than in a).} b) and c) for $^{209}$Pb, again with SkM$^*$.  In
this heavy nucleus we can include orbits with up to only 12 $\hbar\omega$ in
HFODD, and while the mean-field and RPA results for $a_1$ agree very well,
those for $a_0$ differ by about 10\%.  This small discrepancy is almost
certainly due to the limited HFODD model space.  Overall, the level of
agreement, particularly in Ni where we are able to do the best job, convinces
us that both kinds of calculations are essentially correct.

\begin{table}[tb]
\caption{HFODD and RPA results with the Skyrme interaction SkM$^*$ for the
coefficients $a_i$, in $e \, \text{fm}^3$, in the weak-valence-field
approximation (see text) in $^{57}$Ni and $^{209}$Pb.  We have omitted exchange
terms in $V_{PT}$ and taken the zero-range limit of the interaction.  In this
approximation $a_2= 2a_0$. } 
 \label{t:table1}
\begin{tabular*}{.8\columnwidth}{@{\extracolsep{\fill}}llcc} & & $a_0$ & $a_1$
\\
\hline
$^{57}$Ni             & HFODD & -0.0222  & -0.0536 \\
              & RPA   & -0.0226  & -0.0529\\
\hline
$^{209}$Pb            & HFODD & -0.0466  & -0.1059 \\
  & RPA & -0.0507 & -0.1048
\end{tabular*}
\end{table}

\begin{figure}[t]
\includegraphics[width=\columnwidth,clip]{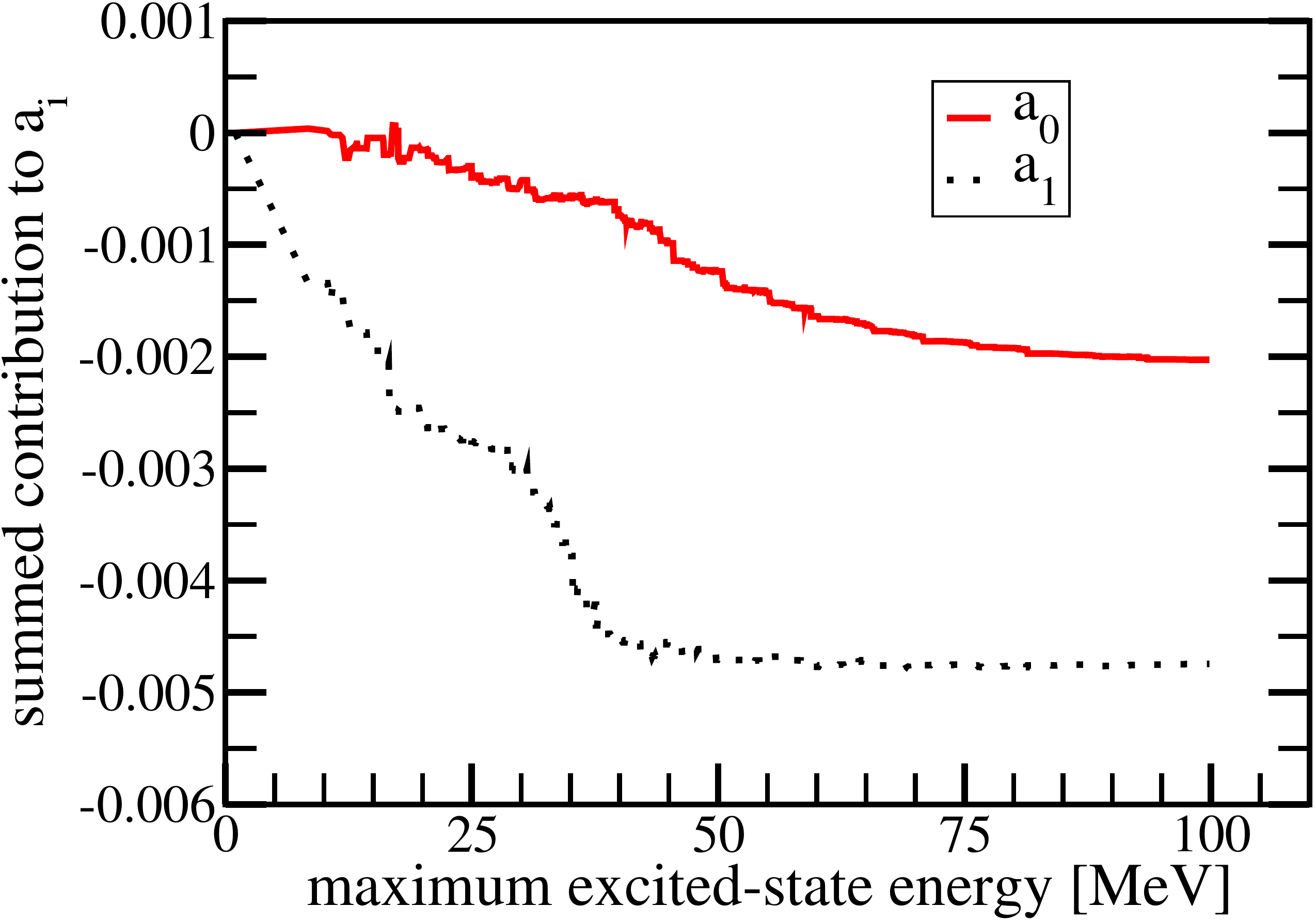}
\caption{\label{f:rpa} (Color online.) The summed RPA contributions to the
$a_i$ in $^{57}$Ni, in the ``weak-valence-interaction'' approximation, as a
function of excited-state energy.}
\end{figure}

The weak-valence-field approximation is equivalent to including only ``diagram
A'' from Ref.\ \cite{jesus05} in the RPA-based diagram sum that yields the
Schiff moment.  We should note that our results for $^{209}$Pb are
significantly different from those for diagram A in the same nucleus given in
the Ph.D.\ dissertation on which Ref.\ \cite{jesus05} was based.  We discuss
possible reasons for the discrepancy, which also exists in $^{199}$Hg, towards
the end of this paper.  For now, we simply note that accurate RPA calculations
require a more careful job than one might think.  Figure \ref{f:rpa} shows the
summed contributions of excited RPA states in Eq.\ (\ref{eq:pert}) to the
$a_i$.  The coefficient $a_1$ is nearly constant after 50 MeV, but $a_0$
continues to decreases even at 80 MeV.  Most RPA calculations do not go that
high in excitation energy, or if they do they make approximations that can
alter results significantly.

\section{Results}
\label{sec:results}

We turn now to the full calculations in nuclei of interest for experiment.  We
apply our mean-field techniques to $^{211}$Ra and $^{199}$Hg.  The first is one
of the Radon isotopes to be explored at TRIUMF \cite{chupp01} and the second is
the nucleus with the best current limit on its Schiff moment.  We use several
Skyrme interactions:  SLy4 \cite{[Cha98]}, SkM$^*$ \cite{[Bar82]}, SV
\cite{[Bei75]}, and SIII \cite{[Bei75]}.  The last of these may not be as
trustworthy as the others; Ref.\ \cite{jesus05} showed that the interaction was
less able to reproduce a related observable, the distribution of isoscalar $E1$
strength in even nuclei.  In previous work we have employed SkO'
\cite{[Rei99]}.  We were not able to find an axially symmetric ground state in
$^{199}$Hg with that interaction, however, and so do not use it here.

The nucleus $^{211}$Ra is spherical, so the calculation there is relatively
straightforward.  We start with an HFB calculation with $H_\text{Skyrme}$
only.  Since the ground state has $J^\pi=\frac{1}{2}^-$, we must block the
lowest $\Omega^\pi = \frac{1}{2}^-$ level, which because of the spherical shape
is essentially the $3p_{1/2}$ orbit ($\Omega$ is the $z$-projection of the
angular momentum in the intrinsic frame).  We then obtain the coefficient $b$
by simply evaluating the expectation value of $S_z^\text{nucleon}/\bar{\rm
g}_0$, with an arbitrary value chosen for $\bar{\rm g}_0$ and $\bar{\rm g}_2$
set to zero.  To obtain the coefficients $a_i$, we follow procedure a) above,
successively setting each of the $\bar{\rm g}_i$ to one.  The results for
several Skyrme interactions appear in Table \ref{t:table2}.
 
\begin{table}[b]
\caption{Results for coefficients $a_i$ and $b$, in $e \ \text{fm}^3$, in
$^{211}$Rn.}
\label{t:table2}
\begin{tabular*}{1.0\columnwidth}{@{\extracolsep{\fill}}lccccc}
&  $a_0$ & $a_1$ & $a_2$ & $b$\\
\hline
SLy4    & 0.042 &-0.018 & 0.071 & 0.016 \\
SkM*    & 0.042 &-0.028 & 0.078 & 0.015 \\
SIII & 0.034 &-0.0004 & 0.064 & 0.015 
\end{tabular*}
\end{table}

The three Skyrme interactions we use give similar results, though the value of
$a_1$ produced by SIII is noticeably suppressed.  The coefficient $b$ is
apparently less sensitive and usually somewhat smaller than the $a_i$.  It is
not small enough, however, to be neglected, as it has been in all prior work.

In $^{199}$Hg the calculation is harder because the nucleus may not be
spherical, and is almost certainly soft.  The energy as a function of
deformation is probably very flat, and the energies of several mean-field
minima may not be very different. For this reason, we do several calculations,
some at deformed minima and some at spherical minima.  Another issue is that
HFODD cannot carry out angular-momentum projection if pairing is included.  We
can estimate the effects of projection, or turn pairing off and carry it out
explicitly.  We follow both courses here and compare the results.  We sometimes
encounter the further problem that the state with the correct ground-state
quantum numbers ($\Omega^\pi = \frac{1}{2}^-$) is not the lowest state in our
calculation.  In a soft nucleus, such an occurrence is not totally surprising.


Finally, the inclusion of $V_{PT}$ causes some problems that are not present
without it.  Although $V_{PT}$ is very weak, the iterative HF energy sometimes
eventually diverges, probably because our axially symmetric excited-state
solution is very slightly unstable against some kinds of asymmetric
deformation.  In such cases, however, the solution converges for awhile, coming
quite close to self consistency, before the weak instability leads it in a
different direction.  We can therefore extract an axially-symmetric result from
the relatively early iterations, during which the solution apparently
converges.  Although we don't have a truly self-consistent solution here, we do
obtain a kind of ``most nearly self-consistent axially-symmetric'' solution,
which is the best we can do without the more difficult and possibly less
meaningful task of considering triaxial shapes for soft systems.

\begin{table}[t!]
\caption{Results for coefficients $a_i$ (in $e \ \text{fm}^3$) in $^{199}$Hg,
with the Skyrme interaction SLy4, in various approximations.  The solution is
axially symmetric with $\beta=-0.13$ and an excitation energy for the
$\Omega^\pi=\frac{1}{2}^-$ state of 0.97 MeV}
\label{t:table3}
\begin{tabular*}{1.0\columnwidth}{@{\extracolsep{\fill}}lccc}
& $a_0$ & $a_1$ & $a_2$ \\
\hline
One HF iteration with $V_{PT}$   & 0.045 &  0.049 & 0.090 \\
Full HF, no projection           & 0.039 & -0.019 & 0.066 \\
Full HF, projected & 0.013 & -0.006 & 0.022
\end{tabular*}
\end{table}

Table \ref{t:table3} displays the results for the interaction SLy4 in
successively better approximations.  The first line shows the results after
including $V_{PT}$ for one Hartree-Fock iteration (starting from the converged
solution with $V_{PT}$ omitted).  In this limit, $V_{PT}$ can excite the core,
but the excited nucleons do not further interact before contributing to the
Schiff moment; that is, no core collectivity is included.  A comparison of the
first two lines shows, in agreement with Refs.\
\cite{dmitriev03,dmitriev05,jesus05}, that collectivity has a large effect on
the $a$'s.  But in contrast to those investigations, we find that collectivity
has a large enough effect on $a_1$ to change its sign.  This change in sign
appears in many of our other Hg calculations as well, even for spherical
minima.  Its appearance there is surprising because the diagrammatic
calculation of Ref.\ \cite{jesus05} used the same Skyrme interactions and
essentially the same spherical-HFB starting point\footnote{One difference is
that the last neutron was in a canonical-basis quasiparticle state in that
work.}, and included much of the same collective physics.  We have already
remarked, though, that where we can check the QRPA results (in $^{209}$Pb) we
do not agree with them.

Another surprising result is that projection reduces the coefficients by a
factor that is very close to three, the same factor as in the rigid-rotor
model.  The reduction factor is nearly three with other Skyrme interactions as
well.  The relatively small deformation of $^{199}$Hg led us to expect a milder
reduction.

\begin{figure}[t]
\includegraphics[width=\columnwidth]{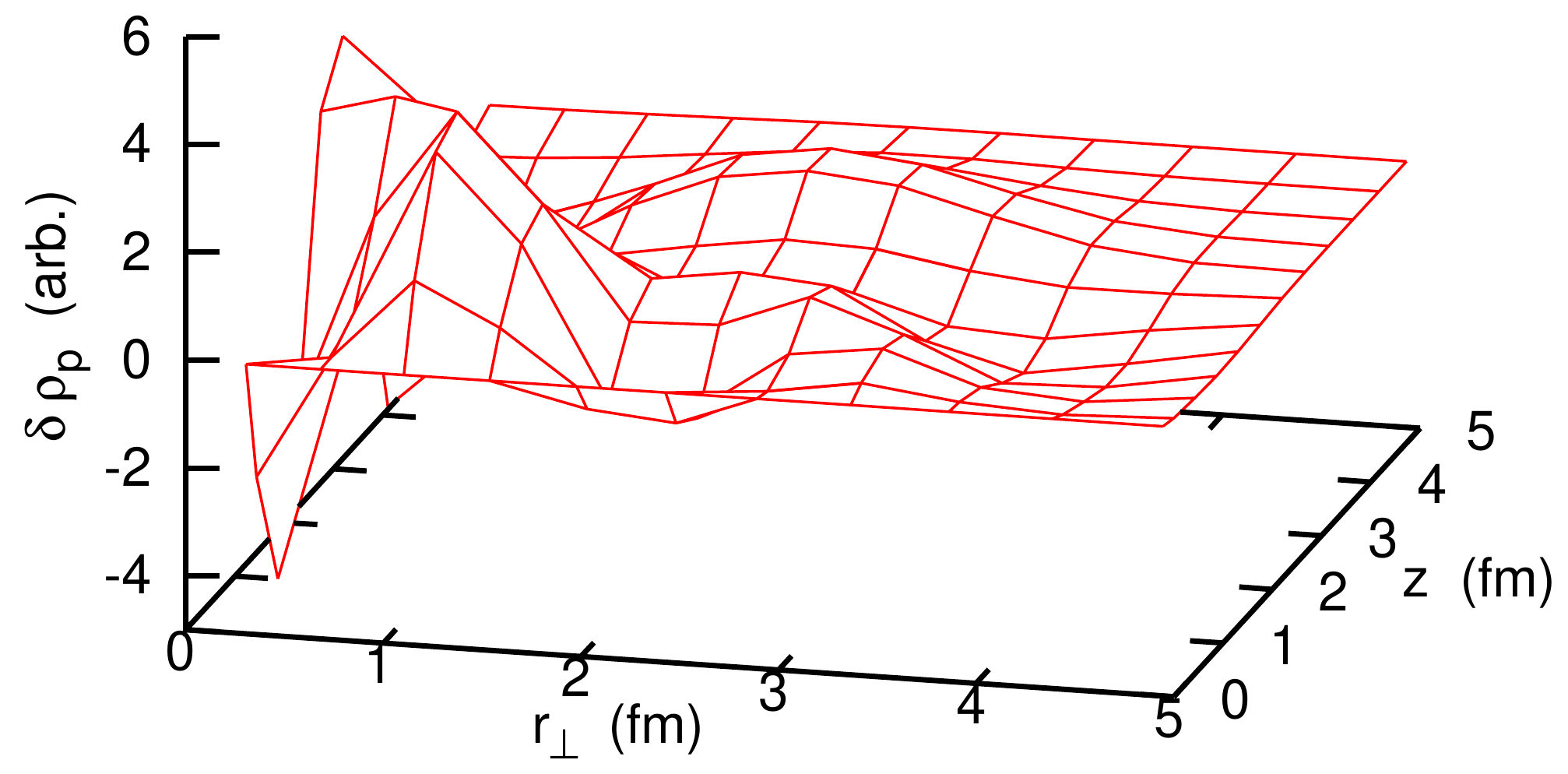}
\caption{\label{f:surface} (Color online.) The change in proton density induced
by the $\bar{\rm g}_1$ term in $V_{PT}$, as a function of $r_\perp \equiv
\sqrt{x^2+y^2}$ and $z$.  The units are arbitrary because of the arbitrariness
in the constant $\bar{\rm g}_1$.  Only $1\!/\!4$ of the nuclear profile is
shown; the density change is symmetric in $r_\perp$ and antisymmetric in
$z$.  The densities were actually evaluated at $13^2$ Gaussian integration
points, a fact that explains the spikiness of the plot.}
\end{figure}

What is unsurprising is that the $a_i$ are delicate and very hard to predict
ahead of time.  Figure \ref{f:surface} shows the change in proton density
$\delta\rho_p$ caused by the inclusion alongside SLy4 of the $\bar{\rm g}_1$
term in $V_{PT}$ (that is, the other $\bar{\rm g}$'s are set to zero).  The
integral of this density difference over $z$ and $r_\perp \equiv
\sqrt{x^2+y^2}$, weighted by $(r^2 -5/3 \braket{r^2}_\textrm{ch})z$, is what
gives the intrinsic Schiff moment (before projection).  The oscillations are
actually even wilder than the figure shows; a deep trough is hidden behind
large peak at small $r_\perp$.  These oscillations make it hard to supply an
explanation for the sign and magnitude of $a_1$.

We turn finally to the full results obtained by using the HO basis of up to 12
$\hbar \omega$ of excitation energy, displayed in Tab.\ \ref{t:table4}.  The
top three lines give the results of deformed HF calculations (the calculation
with SKM* does not give a convergent axially-symmetric result).  All the
interactions under-bind the nucleus; the measured binding energy is 1573.19
MeV.  With the interactions SLy4 and SV, the ground state, as discussed above,
does not have the correct quantum numbers, and we are forced to use an excited
particle-hole configuration that does.  As also mentioned, the energy
eventually begins to diverge from our solution, presumably because of a very
weak triaxial instability.  By contrast, SIII gives the correct ground state,
and no long-term divergence.  After projection, all three calculations produce
similar coefficients $a_0$ and $a_2$, but $a_1$ varies significantly, even in
sign.  We are unable to project the one-body densities that yield the $b$
coefficient, so we take the reduction from the unprojected value to be the same
as that of the $a_i$'s.

\begin{table}[b]
\caption{Results for coefficients $a_i$ and $b$, in $e \, \text{fm}^3$, in
$^{199}$Hg. The third column gives ground-state energy in MeV, the fourth the
deformation, and the fifth the excitation energy (also in MeV) of the lowest
configuration with the same value of $\Omega^\pi$ as the experimental ground
state.  The first three lines are in the HF approximation, while the next two
are in the HFB approximation.  The last two lines report results of previous
work, with the numbers for Ref.\ \cite{jesus05} representing the average over
several interactions.}
\label{t:table4}
\begin{tabular*}{1.0\columnwidth}{@{\extracolsep{\fill}}lccc|cccc}
& $E_\text{gs}$ & $\beta$ & $E_\text{exc.} $ & $a_0$ & $a_1$ & $a_2$ & $b$\\
\hline
SLy4 &-1561.42  & -0.13 & 0.97 & 0.013 &-0.006 & 0.022 & 0.003\\
SIII &-1562.63  & -0.11 & 0    & 0.012 & 0.005 & 0.016 & 0.004\\
 SV   &-1556.43 & -0.11 & 0.68 & 0.009 & -0.0001 & 0.016 & 0.002 \\  
\hline
SLy4 &-1560.21  & -0.10 & 0.83 & 0.013 &-0.006 & 0.024 & 0.007\\
 SkM$^*$ &-1564.03 & 0 &0.82 &0.041 &-0.027 & 0.069 & 0.013\\  
\hline
Ref.\ \cite{dmitriev03} & ---  & --- & --- &0.0004 &0.055 & 0.009 &---\\
Ref.\ \cite{jesus05} & --- & --- & --- &0.007 &0.071 & 0.018 &---
\end{tabular*}
\end{table}

The middle two lines of Tab.\ \ref{t:table4} show the results of HFB
calculations, in which pairing is included.  The SLy4 solution is deformed, and
as mentioned above, we cannot project HFB states; we therefore use the rigid
rotor limit to obtain the projected results in line 4.  SkM$^*$ has a spherical
minimum when pairing is included, so no projection is necessary.  The results
of that calculation are similar to the unprojected results from deformed
solutions.  We conclude that the presence of deformation, at least in our
approach, significantly decreases calculated Schiff moments.

We should note that we do not include the $\mathcal{O}(\alpha^2)$ corrections
to the Schiff moment (generating the ``local dipole moment''
\cite{flambaum02}).  Work in simple models suggests that these corrections to
the $a_i$ are on the order of 25\%, though they could be a larger fraction if
the lowest-order $a_i$ are suppressed.

Our $^{199}$Hg results have some significant differences from those obtained
previously.  Those of the two most comprehensive calculations appear at the
bottom of Tab.\ \ref{t:table4}, with the average of several calculations
presented for Ref.\ \cite{jesus05}.  Our values of $a_0$ and $a_2$ are in
reasonable agreement with those of Ref.\ \cite{jesus05}, but, as already
mentioned, those for $a_1$ are smaller in magnitude and sometimes have the
opposite sign.  Deformation, of course, is one cause, but, as noted above,
there is disagreement even with our spherical calculations.  One source of
difference may be our treatment of core polarization, which is more complete
and self-consistent than that of the earlier papers; the use of a canonical
basis state for the last neutron in Ref.\ \cite{jesus05} may be another.
Finally, the disagreement between our QRPA tests discussed above and those in
the framework of Ref.\ \cite{jesus05} suggest the possibility of an error in
that calculation (which used an early version of QRPAsph that no longer
exists).  Our many tests of the current approach make it unlikely that our
calculations contain outright errors.  The delicacy of $a_1$, is noteworthy,
however, both because of the complicated spatial $PT$-odd density distribution
(see fig.\ \ref{f:surface}) and the sometimes marginally stable convergence to
axially symmetric solutions, a feature that is particularly pronounced for that
coefficient.

How much can we trust the physical approximations underlying our results? The
calculations presented here are unquestionably more sophisticated and inclusive
than any yet attempted, but it may very well be that still more sophistication
is required.  The apparent softness of $^{199}$Hg implies that the true ground
state is best thought of as a superposition of many different mean-field
states, and a generator-coordinate-based approach \cite{ring80} may be required
to adequately represent the mixing.  Though generator-coordinate calculations
are no longer rare, they have not, to our knowledge, been attempted yet in odd
nuclei.  The future of EDM calculations for this kind of nucleus lies in the
generalization of codes like \texttt{HFODD}.  We will need to move beyond
mean-field theory, and ought to expect our current best numbers to be
noticeably revised when we do.


This work was supported in part by the U.S.\ Department of Energy under
Contract No.\ DE-FG02-97ER41019, by the Polish Ministry of Science under
Contract No.~N~N202~328234, and by the Academy of Finland and University of
Jyv\"askyl\"a within the FIDIPRO programme. Shufang Ban would like to
acknowledge partial support from DOE grant DE-FG02-87ER40365.


\end{document}